# Cheap Recovery: A Key to Self-Managing State

Andrew C. Huang and Armando Fox
Stanford University
{ach, fox}@cs.stanford.edu

## Abstract

Cluster hash tables (CHTs) are a key persistent-storage component of many large-scale Internet services due to their high performance and scalability. We show that a correctly-designed CHT can also be as easy to manage as a farm of stateless servers. Specifically, we trade away some consistency to obtain reboot-based recovery that is simple, maintains full data availability, and only has modest impact on performance. This simplifies management in two ways. First, it simplifies failure detection by lowering the cost of acting on false positives, allowing us to use simple but aggressive statistical techniques to quickly detect potential failures and node degradations; even when a false alarm is raised or when rebooting will not fix the problem, attempting recovery by rebooting is relatively non-intrusive to system availability and performance. Second, it allows us to re-cast online repartitioning as failure plus recovery, simplifying dynamic scaling and capacity planning. These properties make it possible for the system to be continuously self-adjusting, a key property of self-managing, autonomic systems.

## 1 The case for cheap recovery

In large-scale Internet services, *cluster hash tables (CHTs)* have emerged as a critical component in the overall state-storage solution (see Figure 1). One primary advantage of a CHT is its ability to scale linearly to achieve high performance [15]. For this reason, single-key-lookup data like Yahoo! user profiles and metadata for Amazon catalog items is stored in CHTs [35, 42]. Another common design pattern involves using a CHT as a base storage layer and placing more complex query logic in the application. Inktomi's search engine accesses several CHTs on each query, the largest of which is a one trillion entry table that maps a word's MD5 hash to a list of document IDs for pages containing that word [5]. In Ninja [16], atomic compare-and-swap is implemented on top of a CHT to increase programming generality. Although databases make up its storage layer, Ebay performs complex queries that involve cross-node joins and foreign-key constraints in the application to achieve greater scalability [17]. A third type of design involves storing semi-persistent session state in a RAM-only CHT [29]. These examples show that certain types of data do not require the full generality of databases and can instead be stored in a CHT for improved scalability and performance.

Not only do CHTs provide scalability and performance, but as this paper shows, CHTs can be designed to simplify two important challenges in persistent-state management. The first challenge is fast, accurate failure detection. In the presence of all types of transient failures, including "fail-stutter" behavior where performance gradually degrades [2], *fast* failure detection is at odds with *accurate* failure detection. Reacting quickly to potential failures leads to false positives, while waiting to collect enough observations to accurately identify failures results in a higher mean-time-to-repair. The second challenge is accurately predicting future load for capacity planning. In a cluster of stateless servers, fluid, reactive scaling avoids having to predict load far in the future [9]. For persistent state, however, the administration and availability cost of repartitioning data makes scaling more expensive and places higher importance on accurate load prediction.

We demonstrate that failure detection and load prediction need not be so accurate if recovery can be made extremely cheap, by which we mean predictably fast and having predictably small impact on system availability and performance. First, cheap recovery lowers the cost of acting on false positives so that effective failure detection is not contingent on accuracy. Second, cheap recovery is the basis for our automatic online repartitioning algorithm, which lowers scaling costs. We apply two design principles for achieving cheap recovery at the expense of consistency, but deliver a consistency model with well-defined guarantees that is appropriate for a large range of CHT-based Internet applications, including those described above.

The main practical benefit of cheap recovery is reduced state management costs. In current systems, the administration costs already dwarf hardware and software costs. With a typical company requiring one ad-ministrator per 1-10 terabytes and data demands growing to the petabyte range, simplifying state management is increasingly important, especially for Internet services, which must deliver content from massive datasets for fractions of a penny per access [14]. Traditional databases can take minutes to recover from a failure, and scaling them up requires administrator intervention and nontrivial downtime [18]. Even in systems that mask failures with failover nodes, when read-one-write-all (ROWA) and primary-secondary replication schemes are used, recovery may involve freezing writes while copying missed updates to the recovering node [30]. These issues are not unique to databases, but exist in CHTs as well [15]; however, this paper shows how CHTs can be designed to avoid these recovery and scaling pitfalls. Our contributions are therefore, as follows:

1. We show that cheap recovery, by lowering the cost of acting on false positives, enables the use of unoptimized anomaly-detection techniques and aggressive restart policies for effective failure detection.

2. We present an online repartitioning algorithm that recasts repartitioning as failure plus recovery, allowing the reuse of existing mechanisms for dynamic provisioning of resources to deal with workload changes and heterogeneous node performance.

3. We identify two design principles that trade consistency for cheap recovery and use them to build DStore (Decoupled Storage), a CHT that can serve as a testbed for measuring the failure handling and resource provisioning benefits of cheap recovery as well as for future work on evaluating the effectiveness of various failure detection techniques.

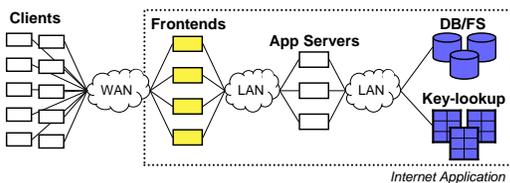

**Figure 1**: Tiered Internet application that uses persistent hash tables as part of its overall storage solution.

In the rest of this paper, we discuss the principles and tradeoffs for achieving cheap recovery (Sections 2-3); we provide implementation details and evaluate recovery behavior (4-5); we describe and evaluate mechanisms for failure detection and repartitioning (6-7); and we conclude by discussing future and related work.

## 2 Two principles for cheap recovery

We follow two design principles for making recovery cheap. The first principle is to tolerate replica inconsistency by using quorum-based replication [13]. In the basic quorum scheme, reads and writes are performed on a majority of the replicas. Since the *read set* and *write set* necessarily intersect, when we use timestamps to compare the values returned on a read, the most up-to-date value is returned. Thus, quorums allow some replicas to store stale data while the system returns up-to-date data. What this means in practice is that a failed replica does not need to execute special-case recovery code to freeze writes and copy missed updates.

Although a wealth of prior work uses quorums to maintain availability under network partitions and Byzantine failures [8, 36][1], few real-life systems do this, perhaps because these failure modes are too rare [43]. Instead, we use quorums to simplify the mechanisms for adding new nodes and rebooting failed nodes, which are frequent occurrences in Internet services. The main cost of quorum-based replication is storage overhead, which we address in the next section.

The second principle is to avoid locking and transactional logging by using single-phase operations for updates. In replicated state stores that use two-phase commit, recovery involves reading the log and completing in-progress transactions by contacting other replicas. Meanwhile, replicas holding data locks for in-progress transactions may be forced to block until recovery is complete to reestablish full data availability. Using single-phase operations, we avoid locking data during failures and cleaning up those locks on recovery. The main cost of single-phase operations is a weaker (but well-defined) consistency model, which we also discuss in the next section.

Our design principles point to two forms of coupling that exist among replicas – the strict consistency among replicas in ROWA and the locking required for two-phase commit. By removing this coupling, we make recovery simpler and less intrusive.

## 3 Cheap recovery tradeoffs

To understand the tradeoffs involved in using quorums with single-phase writes, it helps to have a basic understanding of DStore's architecture (Figure 2).

**Dlibs** (DStore libraries) expose a hash table API and service requests by acting as the *coordinator* for distributed quorum operations on **bricks**, which store persistent data. Based on typical uses of CHTs, we assume the following usage model: When a user issues a request to the Internet service, the request is forwarded to a random application server, which performs one or more hash table operations on the Dlib to fulfill the request[2]. For the discussion on consistency, we consider consistency from the point of view of individual Dlibs making multiple requests as well as from the point of the view of the end-user; however, if a single end-user issues requests from two separate browser windows, we consider each window to be a separate user.

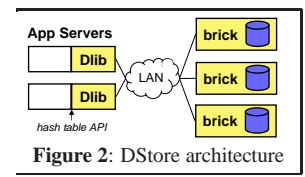

Figure 2: DStore architecture

To handle a request, the Dlib first identifies the bricks responsible for storing the given key (*replica group*). For writes, the Dlib issues the write to all bricks in the replica group, and waits for a majority to respond. As is typical with hash tables, writes completely overwrite the current value. For reads, the Dlib queries a random majority of bricks in the replica group and uses timestamps to determine which value to return. As in Phalanx [32], before returning, if the timestamps do not match, the Dlib issues *read-repair* operations with the up-to-date value-timestamp pair to bricks that returned stale values; this ensures that a majority of the bricks have the up-to-date value upon returning from the read.

---

[1]We reference two survey papers because even a partial list of references would fill entire pages.

[2]Although session affinity is often used to route all of a user's requests to the same application server [22], we do not rely on this mechanism.

## 3.1 Quorums and storage overhead

Quorums require a higher degree of replication than ROWA to achieve an equivalent level of fault tolerance because tolerating $N$ failures requires $N+1$ bricks with ROWA and $2N+1$ bricks with quorums. To capture the entire cost, however, we must consider common failure scenarios and clarify what it means to "tolerate failures." In a cluster, failures are typically either independent or very widespread (as in a site outage) [21]. Under this assumption, tolerating one failure per replica group is sufficient for tolerating most failure scenarios that cluster-based solutions can handle. Furthermore, when considering overall availability, one must take into account availability during recovery. In ROWA, bringing the failed brick up to date causes a big dip in write availability [15]; whereas in quorums with read-repair, the cost of recovery is spread out over time as on-demand repair operations cause a slight dip in write throughput and an increase in read latency for some reads. Therefore, if one needs to meet a certain minimum level of service, using quorums can actually make provisioning for failures simpler and less costly.

Quorums also require a greater number of replica groups to match the read performance of ROWA. On a read, 1 brick is queried in ROWA while $\lceil (N+1)/2 \rceil$ bricks are queried in quorums. To alleviate this, we use a *read-timestamp* optimization in which we read the value-timestamp pair from one replica and read only the timestamp from the remaining replicas. This technique is effective if the value returned is up-to-date and if reading a timestamp is faster than reading the actual value. Since writes are issued to all bricks, the value returned is usually up-to-date, which avoids having to issue a second request to obtain an up-to-date value. Furthermore, when the value size is large compared to the 8-byte timestamp size, most timestamps in the working set can be cached in the brick's RAM. Under these conditions, the overhead of reading timestamps from an in-memory cache is insignificant compared to the cost of reading a value from disk. If, instead, timestamps do not fit in memory, the resulting performance penalty can be resolved by adding more replica groups; however, since administration costs make up a large fraction of the overall storage cost [14], the cost increase is offset by the simplified management cheap recovery provides.

## 3.2 Single-phase writes and consistency

The main challenge in using single-phase operations is ensuring consistency. Two-phase commit guarantees sequential consistency [26], which has two requirements:

1. write atomicity – when a write returns, it has either completely succeeded or completely failed

2. consistent ordering – there is a global ordering of operations that is consistent with the order as seen by individual clients

Distributed consensus, which is necessary for atomicity, has been proven to require at least two phases [31]. Therefore, we aim to guarantee consistent ordering along with a set of well-defined semantics for non-atomic updates. In particular, suppose a write $w$ is issued to replace $v_{orig}$ by $v_{new}$. Any read issued between $w$ and the next attempted write have the following guarantees. If the return status of $w$ is:

- success $\Rightarrow$ reads return $v_{new}$

- failure $\Rightarrow$ reads return $v_{orig}$

- unknown $\Rightarrow$ reads can return $v_{new}$ or $v_{orig}$. If $v_{orig}$ is returned, no user has read $v_{new}$, but a future read by the same or different user might return $v_{new}$. If $v_{new}$ is returned, no user will read $v_{orig}$ in the future

In the rest of this section, we first discuss how we deal with concurrency and failures to provide these consistent ordering guarantees. Then we provide examples of Internet services for which these consistency guarantees support an appropriate usage model.

### 3.2.1 Write concurrency

To handle concurrent writes, Dlibs determine the global update order by generating a globally-unique physical timestamp (local time, IP address) on each update. Following the Thomas Write Rule [40], bricks execute the update only if the new timestamp is more recent than the current timestamp. This way, bricks "agree" on the update order without explicit coordination; however, since timestamps are generated from local clocks, we must be careful to avoid *lost writes*, in which a more recent write is effectively overwritten by a write that occurred in the past:

$U_1$:  $w_1(k,a,ts_1)$
$U_2$:    $a \leftarrow r_1(k)$    $w_2(k,b,ts_2)$    $a \leftarrow r_2(k)$
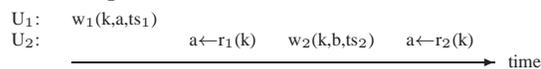

Since $r_2(k)$ returns the previously-wrtten value $a$, it must be the case that $ts_2 < ts_1$ resulting in $w_2(k, b, ts_2)$ being lost; however, $w_2 \rightarrow w_1$ is inconsistent with the order as seen by $U_2$. To prevent inconsistency, bricks return a *timestamp error* for $w_2$ along with the current timestamp $ts_1$. This allows the Dlib for $U_2$ to update its clock, generate a new timestamp, and retry the request.

Synchronizing clocks improves performance by reducing the occurrence of timestamp errors. When clocks are synchronized, in situations where two users issue requests at approximately the same time, it is often acceptable to ignore a lost write without returning an error:

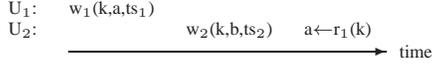

Here, $w_2 \to w_1$ does not violate the order as seen by either user. Furthermore, since we assume that each user's request is handled by a single Dlib acting independently of other Dlibs, $w_1$ and $w_2$ are "unordered" according to Lamport's ordering rules [25]. To take advantage of this, we configure bricks with a small tolerance ($\Delta_{ts}$) and allow a write to be lost without returning an error if the diference in timestamps is smaller than $\Delta_{ts}$. Thus, bricks execute the update if $ts_1 < ts_2$, return a timestamp error if $ts_1 - ts_2 > \Delta_{ts}$, and otherwise, disregard the write and do not return an error. Since ordering issues arise only when a write follows a read, we set $\Delta_{ts} = 1ms$, which is based on the minimum network roundtrip time. Combined with NTP [33] to synchronize clocks to within 1 millisecond, this eliminates almost all timestamp errors; however, even if clocks cannot be synchronized, correctness is not compromised.

#### 3.2.2 Read-write concurrency

If user $U_1$ updates a value while another user $U_2$ issues a read, $U_2$ may witness a *partial write* where the write has reached some, but not a majority of the bricks. When this occurs, depending on which bricks are queried on a read, different values are returned, which can cause ordering inconsistencies like the one shown in Figure 3. In the figure, the unlabelled timeline next to the user's timeline represents a random Dlib that may be different for each user request.

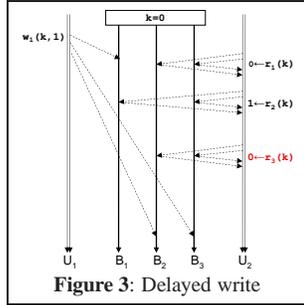

Figure 3: Delayed write

Figure 4 shows how the read-repair mechanism described earlier resolves the partial write problem by synchronously committing the new value before returning from $r_2(k)$. Once committed, all future reads return the new value. It follows that reads issued prior to the commit point returned the old value; otherwise, the new value would have been committed already. Therefore, forcing a commit point using read-repair resolves ordering inconsistencies that can arise from concurrent reads and writes.

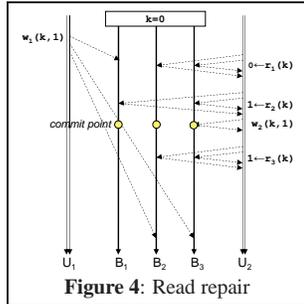

Figure 4: Read repair

#### 3.2.3 Dlib failures

With quorums, brick failures do not affect overall consistency, but when Dlibs fail, partial writes can occur. Like with the read-write concurrency example, read repair resolves partial writes due to Dlib failures (Figure 5). Analogous to recovery in ROWA versus quorums, recovery in two phase versus single phase is a tradeoff between bulk and incremental recovery. In two-phase commit, locked data can cause transactions to block and data to be unavailable for reads and writes when a Dlib fails between phases. In DStore, data remains available, and performance is slightly lower after recovery as read-repair operations resolve partial writes.

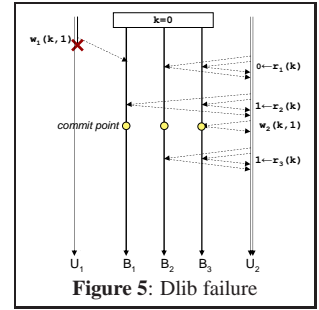

Figure 5: Dlib failure

Under a fail-stop model, quorums with read-repair guarantees linearizability [19], which is stronger than sequential consistency. One key to the proof [36] is that a partial write can be serialized any time after a failure because the coordinator does not recover and no other party knows when the write was actually issued. In DStore, however, due to the assumed usage model, the "coordinator" is not only the Dlib, but also includes the user who issued the request. For this extended view of the coordinator, the fail-stop assumption does not hold. Therefore, we next consider how to provide consistent ordering for the user that issues an update whose status is "unknown."

When a Dlib or the application server it resides on fails, the Web server tier may retry the request on a different application server, or an application-level error may cause the user $U_1$ to resubmit the request via HTTP Retry-After. In either case, any partial write that occurred is overwritten and $U_1$ sees a consistent order; however, if $U_1$ performs a read to check the value, the old value may be returned for awhile before the new value is committed (Figure 6). Unlike $U_2$, $U_1$ knows when $w_1$ was issued, so it does not make sense for $U_1$ to see the update being committed at a later time. This is where DStore's update semantics violate atomicity; on a Dlib failure, the update may have succeeded, failed, or may take effect at some later point in time.

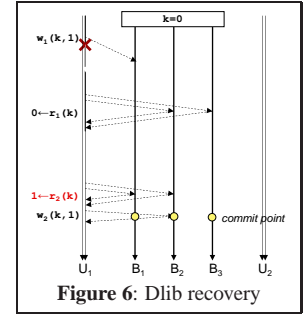

Figure 6: Dlib recovery

To ensure that $U_1$ sees a committed value after a partial write, we must record the fact that a write is issued but has not completed. Rather than adding another phase to the write protocol to keep a write history on the bricks, we effectively store the "state" associated with two-phase commit at the client. When $U_1$ clicks *submit*, client-side JavaScript code writes an *in-progress* cookie on the client before the request is submitted. Upon success, the server returns a replacement cookie with the in-progress flag cleared. On a subsequent read, the cookie is sent along with $U_1$'s request. If cookie in-progress flag is not cleared, the Dlib detects this and reads the values from all bricks to find the most recent value (Figure 7); that value is then written back to a majority of the bricks to commit any partial writes, reestablishing the quorum invariant. As an alternative to using the write-in-progress cookie, Internet services can implement other application-level techniques to force a user to reissue the request before reading.

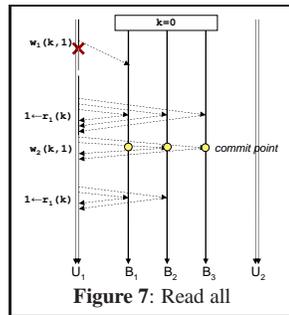

**Figure 7**: Read all

### 3.2.4 Two-phase commit revisited

The two techniques, quourms and single-phase updates, are orthogonal. For example, one could use quorums as the replication scheme, but use two-phase commit to provide consistency. In DStore, we use single-phase operations because our aim is to explore how cheap we can make recovery to discover the resulting properties. The following table summarizes the tradeoffs between using two-phase commit and single-phase operations in DStore:

| Property | 2-phase commit | Single-phase operations |
|---|---|---|
| Consistency | Sequential consistency | Consistent ordering |
| Recovery | Read log to complete in-progress transactions | No special-case recovery |
| Availability | Locking may cause requests to block during failures | No locking |
| Performance | 2 synchronous log writes 2 roundtrips | 1 synchronous update 1 roundtrip |
| Other costs | None | Read repair causes slight performance degradation Relies on client to store write-in-progress cookie |

### 3.2.5 Consistency guarantees

To conclude the discussion on trading consistency for single-phase operations, we summarize DStore's consistency model. DStore enforces a total ordering consistent with the partial order seen by individual Dlibs and end users. More specifically, consider a user $U_1$ who performs a write $w_1(k, v_{new})$ on a hash table in which $(k, v_{old})$ is a key-value pair. Assuming there are no subsequent updates, reads return $v_{new}$ if $w_1$ returns success, and $v_{old}$ if $w_1$ returns failure. If the return status of $w_1$ is unknown, the read guarantees differ for different users. If $U_1$ issues the read, the write-in-progress cookie accompanies the request and the Dlib ensures that a partial write is immediately committed so that the value is returned is seen by all future reads. If another user $U_2$ issues the read, returning $v_{old}$ guarantees that no user has read $v_{new}$, and returning $v_{new}$ guarantees that no user will read $v_{old}$ in the future.

To make the consistency discussion more concrete, we list some applications for which DStore, and its current consistency model, would and would not be appropriate:

- Single-user data *(yes)*: user profiles, shopping carts, workflow data for tax returns

- Single-producer multi-consumer *(yes)*: catalog metadata, search engine data (retailers/crawlers make updates, which are reflected on the site)

- Multi-producer multi-consumer *(probable)*: auction bids (users submit bids, which other users see after some delay) – requires application-level checks that may require atomic compare-and-swap [16]

- Non-overwriting objects *(no)*: workflow data for insurance claims (multiple users update separate portions of a single document) – since updating stale data on a quorum replica has undefined results use ROWA instead

To summarize the tradeoffs described in this section, for a small amount of storage overhead, one can use quorums to simplify recovery and keep data available throughout recovery. On top of that, for many Internet applications for which the consistent ordering guarantees described in this section are appropriate, one can use single-phase operations to further simplify recovery and keep all data available during a failure.

## 4 Implementation details

In this section, we discuss details of the DStore implementation. Recall that DStore is composed of two components, Dlibs and bricks shown in Figure 2.

A Dlib is a Java class that presents a single-system image with the consistency model detailed in the previous section. The Dlib API exposes `put(key,value)` and `get(key)` methods where keys are 32-bit integers and values are byte arrays. Dlibs service requests by issuing read/write requests to bricks via TCP/IP socket connections. In order to act as the *coordinator* for these distributed operations, Dlibs maintain soft state metadata about how data is partitioned and replicated. Finally, Dlibs maintain request latency statistics, which are used in making repartitioning decisions.

Bricks store persistent data accessed via the brick API – `write(key,value,ts)`, `read_val(key)`, and `read_ts(key)`. For reading data, `read_val` returns a value-timestamp pair, while `read_ts` returns only the timestamp. On a `write`, the brick checksums the key-value-timestamp object and writes it synchronously to disk. Bricks also cache timestamps in an in-memory Java Hashtable, and use the file system buffer cache to cache values. Since the timestamp cache is updated after the value is written to disk, if a brick fails while processing a write, on a read, it conservatively returns a timestamp that is no newer than the timestamp of the value on disk.

Currently, data is stored in files as fixed-length records where the record size is configured at table-creation time. Although this scheme is constrained by its fixed-length nature, it is appropriate for datasets with low value-size variance, such as Amazon's catalog metadata database, which has fixed-size entries [42]. Since the underlying storage scheme is orthogonal to DStore's design techniques and is not critical for evaluating the system's recovery and manageability, simplicity is the main reason this scheme is used. Nevertheless, bricks are designed so that it is easy to plug in different storage schemes. For example, implementing wrapper code for Berkeley DB [3] took less than an hour and changing storage schemes takes a matter of seconds.

Bricks maintain two open TCP/IP socket connections (send and receive) per Dlib, one control channel used for initiating recovery and repartitioning, and three message queues (read, put, and ts) serviced by individual thread pools. Like router QoS queues, differentiating requests enables administrators to provision resources and maintain a minimum level of service for each request type. Furthermore, differentiating longer-running write requests from read requests reduces the service time variance, and subsequently, the average queuing delay [41].

### 4.1 Keyspace partitioning

Storage and throughput capacity is scaled by horizontally partitioning the hash table across bricks. Each partition is replicated on a set of bricks to form a *replica group*, the unit upon which quorum operations are performed. Keys are partitioned across replica groups based on the keys' least-significant bits, called the *replica group ID* (*RGID*). Later we show how the keyspace can be dynamically repartitioned without loss of availability to compensate for an uneven workload or heterogeneous brick performance.

Upon startup, each brick is configured with an RGID-mask pair, which it beacons periodically to distribute metadata and indicate liveness. For example, if brick $B$ beacons $(1, 11)$, this tells Dlib's to use the last two bits of the key to find $B$'s RGID, $01$. Using this information, Dlibs build a soft-state RGID-to-brick mapping (*RGID map*); to find the replica group storing a given key, Dlibs find the entry with the longest matching suffix.

The beaconing period is a system-configurable parameter (currently set to two seconds), so in between beacons, a Dlib's RGID map may become stale. Since each brick is the ultimate authority of its own RGID, if a Dlib sends a request to the wrong brick, the brick returns a `WRONG_REPLICA_GROUP` error. Once the Dlib's RGID map is updated on the next beaconing period, the request is sent to the correct set of bricks. Finally, a brick can be a part of more than one replica group by announcing multiple RGIDs. This feature is used to spread load across a heterogeneous set of bricks and in the online repartitioning algorithm described later in this section.

### 4.2 Detailed algorithms

DStore's `put` and `get` algorithms, which were outlined earlier, are described here in full detail. For this discussion, let $N$ be the number of bricks in a replica group, let $WT$ (*write threshold*) be the minimum number of bricks that must reply before a `put` returns, and let $RT$ (*read threshold*) be the minimum number of bricks that are queried on a `get`. In DStore, we choose $WT$ and $RT$ to be a majority – $\lceil (N+1)/2 \rceil$; however, in general, $WT$ and $RT$ can be chosen such that the read and write sets intersect: $WT + RT > N$.[3]

**Dlib put**: On a `put`, the Dlib generates a physical timestamp from its local clock and appends its IP address. The timestamped value is sent to the entire replica group, but the Dlib only waits for the first $WT$ responses to ensure the quorum majority invariant holds, and ignores any subsequent responses. Pseudocode for all algorithms are provided in the appendix.

**Brick write**: When a brick receives a write request, it overwrites a value only if the new value has a more recent timestamp. If the new timestamp is older than the current timestamp by $\Delta_{ts}$, which is set to 1 millisecond, the brick returns `TIMESTAMP_ERROR`.

**Dlib get**: On a `get`, the Dlib selects $RT$ random bricks from the replica group, and issues a `read_val` request to one brick and `read_ts` requests to the remaining $RT - 1$ bricks. After the value and all the timestamps are returned, the Dlib calls the `check` method to confirm that the value is up to date.

**Dlib check**: On each `get`, two checks are performed on the timestamps. First, the Dlib checks whether the timestamp for `value` is the most recent one returned. If it is, `value` is returned; otherwise, the value is read from the brick that returned the most recent timestamp. Second, the Dlib checks whether at least $WT$ bricks have the most recent timestamp. If not, the most recent value and timestamp are written back to ensure that enough bricks contain an up-to-date value. This repair mechanism is used to repair partial writes arising from Dlib failures.

---
[3]Quorum systems generally also require that $2 * WT > N$ so that simultaneous writes can be ordered; however, we use physical timestamps, which eliminates this requirement.

## 4.3 Restart mechanism

The final implementation detail we discuss is the brick recovery mechanism, which is used by the failure detection mechanisms described later. To restart a failed brick $B_f$, a Dlib scans its RGID map for the brick with the next-highest IP address $B_r$, and sends a RESTART_BRICK message to $B_r$'s control channel. If $B_r$ does not respond because it too has failed, the brick with the next-highest IP address is asked to restart both bricks. The Dlib also removes $B_f$ from its RGID map so that requests are not sent to $B_f$ until it recovers.

To restart the brick, $B_r$ runs a script that first sends a `kill -9` to $B_f$'s brick processes. Next, the script attempts to restart the brick processes. If instead, $B_f$'s machine does not respond at all, the node can be restarted using an IP-addressable power source. Since multiple Dlibs are likely to detect $B_f$'s failure, $B_r$ does not perform recovery more than once every four seconds (two times the beaconing interval). Further, since restarting a brick only cures transient failures, if the brick has been restarted more than a threshold number of times in a given period, it is assumed the brick has a persistent fault and should be taken offline. Automatically reconstructing data after disk failures and allowing failed components to be replaced en-masse, say on a weekly basis, are next steps in our future work.

## 5 Recovery benchmarks

In this section, we evaluate how well DStore achieves its goal of cheap recovery. We first measure basic performance and scalability, and then show system behavior during failure and recovery.

### 5.1 Benchmark details

All benchmarks are run on the UC Berkeley Millennium cluster, which has 42 PCs with dual 1GHz Pentium III CPUs, 1.5GB RAM, and dual 36GB IBM UltraStar 36LZX hard drives. Nodes are connected by a 100Mb/s switched Ethernet. A single instance of a brick or client application is run on each node using Sun's JDK 1.4.1 on top of Linux 2.4.18. DStore is configured with three bricks per replica group ($N = 3$) and 1-KByte records. The number of threads allocated to service the write, read_val, and read_ts queues is 14, 32, and 10 respectively, which produces a workload mix of around 2.5 to 5 percent writes. To generate load, clients perform closed-loop operations[4] on random keys and 1-KByte values;

[4]closed-loop = a new request is immediately issued after the previous request returns

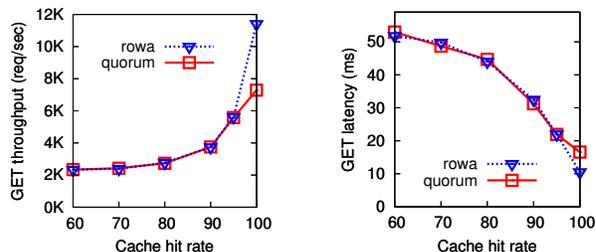

**Figure 8**: **ROWA vs. Quorums:** Under realistic cache hit rates, reading extra timestamps from an in-memory in quorums has little effect on performance because the disk is the bottleneck.

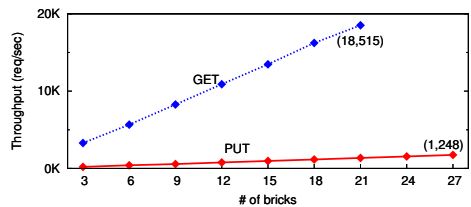

**Figure 9**: **Linear scaling:** GET and PUT throughput scale linearly with the number of bricks (cache hit rate = 85%).

this 1-KByte value size falls within the range of typical hash table object sizes in Internet services [35]. Enough clients are run to saturate the bricks in steady-state and when there are concurrent reads and writes, the ratio of read clients to write clients is 4:1; however, as shown below, the GET and PUT throughput capacities are independent of the workload mix and are instead determined by the number of threads allocated to each request queue.

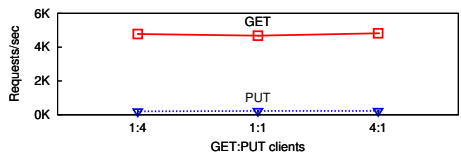

### 5.2 Steady-state performance

Although designed for manageability, DStore retains the high performance and scalability of CHTs. First, we show that under reasonable cache hit rates, the read-timestamp optimization gives quorums read performance comparable to ROWA. Second, we show that DStore throughput scales linearly with the number of bricks.

**Timestamp read overhead.** To evaluate the effectiveness of DStore's read-timestamp optimization, we compare the read performance of ROWA versus quorums under different cache hit rates. We run a three-brick DStore with 2GB of data and induce approximate cache hit rates between 60 and 100 percent. The client induces a hit rate $c$ by generating a random integer $i$ between 1 and 100 on each request. If $i \leq c$, the key $k$ is chosen to fall within the 1.2GB working set: $0 \leq k < 1.2M$. If $i > c$, $k$ is chosen to induce a disk access: $1.2M \leq k < 2M$.

Figure 8 shows that although ROWA outperforms

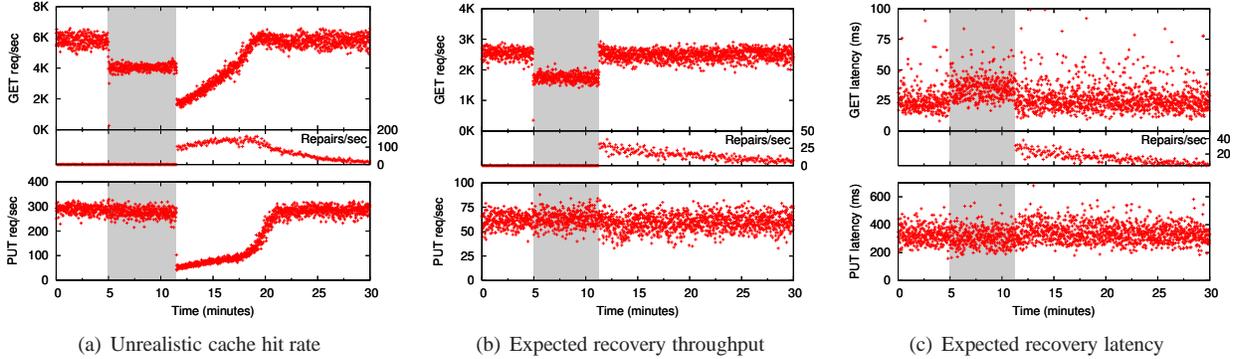

(a) Unrealistic cache hit rate    (b) Expected recovery throughput    (c) Expected recovery latency

**Figure 10**: **Recovery behavior:** In (a), under a 100% cache hit rate, recovery is neither fast nor non-intrusive because of the disparity between disk-bound recovery performance and in-core steady-state performance. In (b) and (c), under a realistic cache hit rate of 85%, the extra disk accesses from a small number of read repair operations does not greatly affect throughput or latency because the steady-state performance is already disk-bound.

DStore when the working set fits in memory, the disk quickly becomes the bottleneck for lower, more realistic cache hit rates and performing the extra timestamp read becomes largely insignificant. If timestamps do not completely fit in memory, the 100-percent cache hit rate measurements can be seen as the worst-case overhead.

**Linear performance scaling.** To evaluate DStore's scaling properties, we induce an 85-percent cache hit rate as described above and measure GET and PUT throughput separately. Figure 9 shows throughput scaling linearly up to 21 bricks for GETs and 27 bricks for PUTs after which, we did not have enough nodes to scale the benchmark further.

## 5.3 Fast, non-intrusive recovery

Next, we show that recovery is fast and leaves data available for reads and writes throughout failure and recovery. For these benchmarks, we run a three-brick DStore, induce a brick failure at $t = 5min$, and manually restart the brick at $t = 10min$. For these and all subsequent benchmarks, we show the get/put throughput and the repair operations per second. The shaded areas highlight the period a brick is offline, either due to a failure, reboot, or to intentionally stop receiving requests like during repartitioning. Each point in the graph represents a single throughput measurement (taken once per second) as seen from the clients.

Figure 10(a) shows recovery behavior under a 100-percent cache hit rate. As expected, get throughput drops by one third during failure and put throughput remains steady because writes are issued to the entire replica group causing bricks to see roughly the same load no matter how many bricks are available. After the brick recovers, get throughput drops dramatically because of the requests that require repair. Although the percentage of repairs is small, the high cost of disk writes causes reads to become disk-bound. Since repair operations are placed in the bricks' write queues, contention causes put throughput to drop as well. Throughput returns to normal only after about ten minutes. The dramatic drop in throughput capacity that takes ten minutes to restore can hardly be considered fast and non-intrusive.

According to an industry expert, workloads for large data sets are typically disk-bound with cache hit rates ranging from 60 to 90 percent [35]. As shown in Figures 10(b) and (c), recovery behavior changes significantly when the workload induces a more realistic cache hit rate of 85%. As in the previous benchmark, throughput drops during failure for get requests, but not put requests. In fact, put throughput rises slightly due to the slight drop in contention from get requests requiring disk access. On recovery, get throughput immediately returns to normal because the small number of repair operations has little effect on the already disk-bound workload. put throughput drops due to contention from repair requests; however, the effect is less pronounced because in steady-state, put and get requests already contend for the disk. In 10(c), as expected, get latency increases when the brick fails because the same load is being handled by fewer bricks. On recovery, put latency increases due to an increased write load from repair operations. In general, the longer the brick failure, the more read-repair operations are required and the more contention there will be on the disk for put requests. Whereas in ROWA where writes are frozen while data is copied, this benchmark shows that the read-repair mechanism spreads the cost of recovery over time with modest performance impact.

## 6 Simple, aggressive failure detection

Two failure detection mechanisms can cause DStore to initiate brick recovery as described earlier in Section 4. First, Dlibs initiate recovery if a brick misses two consecutive RGID beaconing periods. This simple beaconing mechanism is usually sufficient to detect stopping failures like node crashes, or faults that can be mapped to stopping failures using fault-model enforcement, like certain types of memory corruption and bit flips [34].

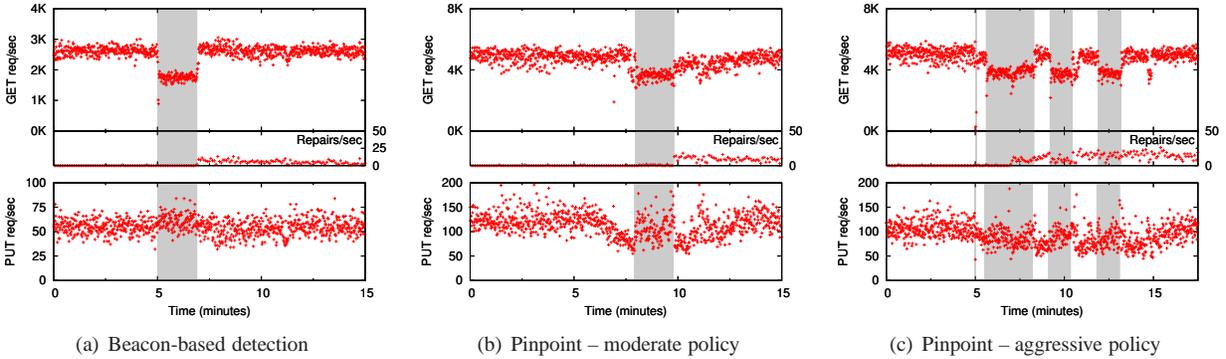

**Figure 11**: **Fast failure detection:** In each benchmark, failure or degradation is induced at $t = 5min$. In (a), beacon-based detection immediately detects the brick failure. In (b), when $anomaly threshold = 8$, Pinpoint detects brick degradation after a slight dip in overall system performance. In (c), when $anomaly threshold = 5$, Pinpoint detects the degraded brick earlier and even though it also raises false positives, due to the low cost of recovery overall performance is not seriously affected.

Whereas stopping failures are relatively easy to detect, fail-stutter is difficult to detect reliably because it is hard to determine if degradation is due to faulty component or some other factor like garbage collection or cache warming. To detect fail-stutter behavior and other anomalous behaviors that may be indicative of a current or pending failure, a brick periodically reports operating statistics, which are compared to the operating statistics of other bricks to detect deviant behavior. If a brick's operating statistic deviates by more than a certain deviation threshold, an anomaly is raised. If the number of anomalies for one brick exceeds a certain anomaly threshold, the brick is restarted. The set of operating statistics bricks report is listed below.

| Statistic | Description |
|---|---|
| CPU load | Load average as reported by `uptime` |
| Memory usage | Memory usage as reported by `free` |
| Queue length | Current GET, PUT, TS queue lengths |
| Success | GET, PUT, TS requests processed since last report |
| Dropped | GET, PUT, TS requests dropped since last report |
| Queue delay | Queuing delay EMA (exponential moving average, $\alpha = .0002$) for each queue type |
| Latency | Request processing time EMA ($\alpha = .0002$) for each request type |

We use two statistical methods for generating a model of "good behavior" and detecting anomalies. The first method is median absolute deviation, which compares one brick's current behavior with that of the majority of the system. This metric was chosen for its robustness to outliers in small populations, which is important for small DStore installations. The second method is the Tarzan algorithm [23] for analyzing time series, which incorporates a brick's past behavior and compares it with that of the rest of the bricks. For every statistic of each brick, we keep an N-length history, or time-series, of the state and discretize it into a binary string. To discover anomalies, Tarzan counts the relative frequencies of all substrings shorter than k within these binary strings. If a brick's discretized timeseries has an abnormally high or low frequency of some substring as compared to the other bricks, an anomaly is raised. We set $k = 3$, $N = 100$, deviation threshold to $0.5$, and we vary the anomaly threshold.

We note that although such methods are potentially powerful tools for identifying deviant behavior, identifying the "best" algorithms to use is a *non-goal* of this work. Rather, our goal is to demonstrate that DStore's cheap recovery allows us to take advantage of these relatively simple, application-generic techniques: even though anomalous conditions may be false positives that predict an eventual brick failure, the low cost of recovery enables us to act on some false positives without serious adverse affects. In fact, at times, it may be beneficial to reboot a brick even if an anomaly is transient.

### 6.1 Failure detection benchmarks

These benchmarks show how DStore takes advantage of cheap recovery to detect stopping and slow-down failures using aggressive failure detection. For these, and all future benchmarks, we induce an 85-percent cache hit rate. Also, since latency results are qualitatively very similar to those shown in Figure 10c, for all remaining benchmarks, we show only throughput graphs.

In the first benchmark, we run a three-brick DStore and kill one brick at $t = 5min$. Figure 11(a) shows that DStore's beacon-based detection detects the stopping failure and restores full capacity within about a minute. To evaluate DStore's statistical anomaly-detection mechanisms, in the next two benchmarks, we run a six-brick DStore and at $t = 5min$, gradually degrade one brick's throughput capacity by increasing its request processing latency. We model degradation as lost CPU cycles, which is reasonable for slow downs due to memory leaks and virtual memory thrashing. We simulate degradation by performing extraneous floating point operations before processing each request, the number of which is increased every ten seconds.

Figures 11(b) and 11(c) show failure detection with varying degrees of aggressiveness. Recall that the anomaly threshold corresponds to the number of brick statistics that must indicate deviant behavior before a brick is rebooted. By trial-and-error, we selected two thresholds, one that showed some system degradation before recovery was initiated, and another that caused spurious reboots in bricks in which we did not inject faults. With an anomaly threshold of 8, DStore detects the injected slow-down failure and reboots the node after a small, but noticeable degradation in system performance. With a more aggressive policy where the threshold is set to 5, the degrading brick is caught more quickly, but DStore also reboots other bricks in which no faults injected; however, as discussed before, the low cost of acting on false positives allows us to use aggressive failure detection without worrying about the extra reboots.

These benchmarks show that with cheap recovery, effective failure detection does not require the best algorithms with highly-tuned parameters that reliably detect failures without raising false positives. Instead, aggressive anomaly-detection can be used in DStore for effective, low-cost failure handling.

## 6.2 Remarks on rolling reboots

Along with beacon-based and statistical anomaly-based failure detection, cheap recovery enables a third, complementary failure handling mechanism – rolling reboots. By proactively rebooting bricks, software rejuvenation [20] can prevent failures that arise due to aging effects like memory leaks. Although rolling reboots catch a superset of the problems we can detect using beacons and statistical anomaly-detection, DStore detects and deals with failures more quickly than if we had to wait for the reboot to cycle to the affected brick. Thus, all three are complementary mechanisms.

## 7 Zero-downtime incremental scaling

Cheap recovery is the basis for our online repartitioning algorithm. Currently, repartitioning is initiated when an administrator issues a command with the new bricks' hostnames to add the new bricks; however, DStore is amenable to systems that monitor workload and predict resource utilization to automatically decide when to bring more resources online [27]. Once the new bricks are added, deciding *which* bricks to repartition and integrating the new bricks into the system is handled automatically.

## 7.1 Repartitioning algorithm

The steps in our automatic, online repartitioning algorithm are as follows:

1. Discover replica group information. The new brick, $B_{new}$ constructs an RGID map by listening to RGID beacons.

2. Select the brick to repartition. $B_{new}$ listens for Dlib latency beacons and selects the repartition brick $B_r$ based on average request latency.

3. Split $B_r$'s replica group. $B_{new}$ sends a SPLIT_RG command to the control channel of each brick in $B_r$'s replica group. A brick logically splits its RGID by adding an extra bit and announcing two RGID's; for example, if $B_r$'s current RGID is 0, it begins announcing both 00 and 10. $B_{new}$ listens for updated RGID beacons from $B_r$'s replica group to make sure all bricks have split before continuing.

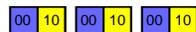

4. Take $B_r$ offline. $B_{new}$ sends an OFFLINE command to $B_r$. $B_r$ attaches an "offline" flag to its RGID beacons causing Dlibs to stop sending requests to $B_r$. From the point of view of the Dlibs, it appears as if $B_r$ has failed; the only difference is that the Dlibs do not initiate recovery.

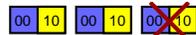

5. Copy data from $B_r$ to $B_{new}$.

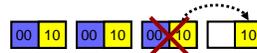

6. Set $B_r$'s and $B_{new}$'s RGIDs via their control channels, and bring both online. RGID's are set so that the partition is physically split: $RGID(B_r) = 00$ and $RGID(B_{new}) = 10$.

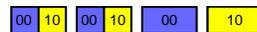

First, consider the algorithm's online aspect. Taking $B_r$ offline and bringing it back online has the same effect as if $B_r$ had failed and recovered; the only difference is that upon "recovery," two physical bricks have taken $B_r$'s place. Therefore, any updates that occur during repartitioning are executed by the online bricks with updates being propagated to the offline bricks via the read-repair mechanism. Since repartitioning has the same effect as a failure, when multiple bricks are added to the same replica group, they are integrated into the system one at a time. If a failure occurs while repartitioning and the number of online bricks falls below a majority, the repartitioning process is halted and $B_r$ is brought back online. The resulting effect is no different than if the $B_r$ had simply failed and recovered. Although it is possible to add bricks simultaneously without taking bricks offline by adjusting the read and write thresholds accordingly, we use this recovery-based mechanism because it makes the performance impact and time for adding new bricks more predictable.

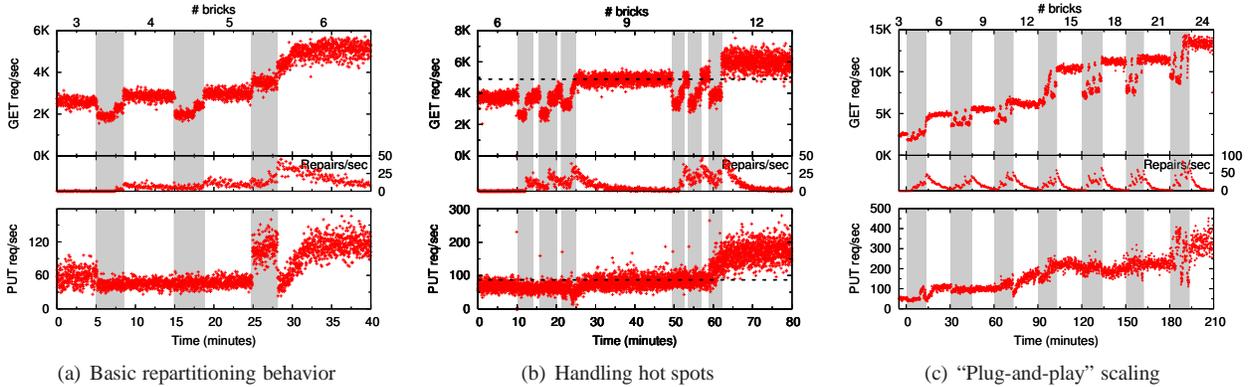

**Figure 12**: **Online repartitioning:** In (a), three bricks are added to a three-brick DStore to double throughput without causing unavailability. In (b), six bricks are added to a six-brick DStore in which clients induce a data hotspot in keys ending in 00 to show that the latency-based scheme for making partitioning decisions produces higher performance than a naive approach, which is shown by the dashed line. In (c), repartitioning is used to scales DStore from 3 to 24 bricks in an automatic, fully-online fashion.

Second, consider the automatic aspect of the algorithm. DStore's simple hash table API removes implicit data dependences, which enables our algorithm to select $B_r$ based solely on load information without worrying about splitting up data that is accessed together. The generalized form of the algorithm is to globally repartition the system by selecting the $N$ most loaded bricks and moving portions of the keyspace to the new brick as necessary; however, for simplicity, the current algorithm splits only the heaviest-loaded brick, approximated by the exponential-moving-average request latency with a smoothing constant $\alpha = 0.5$.

## 7.2 Repartitioning benchmarks

These benchmarks show that repartitioning, like recovery, has a predictably small impact on availability and performance. We first show basic repartitioning behavior by running a three-brick DStore and doubling the number of bricks by adding a brick every ten minutes. In Figure 12(a), the first two shaded areas are similar to brick failure and recovery. The slight throughput increase at the end of the gray area results from the original brick coming back online before the new brick preloads its timestamp table and starts up (analogous to cache warming). After two bricks are added, the throughput increase is slight because capacity is limited by the non-partitioned brick. Load-based request distribution would enable DStore to take advantage of the spare capacity of the repartitioned portions of the replica group by biasing where requests are sent. While the third brick is repartitioned, throughput rises because the rate-limiting brick is removed, leaving the system with two replica groups each with two bricks. After the third brick is repartitioned, throughput climbs up to a new steady-state level, approximately double the original level.

The second benchmark shows that the latency-based brick-selection scheme handles hotspots by repartitioning the most heavily-loaded bricks. In Figure 12(b), we start with a six-brick DStore and two replica groups, $RGID = 0$ and $RGID = 1$. Rather than spreading requesting evenly across the keyspace, clients request keys in a biased fashion, producing a data hotspot in keys ending in 00. Three bricks are added at $t = 10m$ to split the 0 partition. This addition brings the throughput to the same level as a naive partitioning in which the replica groups are split evenly across the entire keyspace while the 00 remained heavily overloaded (represented by the dashed line). Instead, when the second set of bricks is added at $t = 50m$, DStore splits the 00 partition into 000 and 100 to achieve much higher performance.

In the final benchmark, we use online repartitioning to scale DStore up to 24 bricks. In Figure 12(c), going from three to six bricks doubles throughput; however, throughput does not double as expected, at 12 bricks. Observing the brick statistics collected for failure detection shows some bricks consistently exhibiting relatively poor performance despite equivalent hardware/software configurations. As described in [1], performance variations can arise in a homogenous cluster due to differences in disk layout and memory management; this effect is particularly acute in I/O-bound workloads. As is the case when repartitioning a single replica group, the throughput increase follows a step function rather than a linear one. This is due to the fact that before the number of bricks is doubled, only part of the keyspace has been partitioned while requests remain evenly distributed across the keyspace. Global repartitioning would produce a greater throughput increase for each incremental brick addition; however, unlike the evenly-distributed workload of this benchmark, we expect real workloads to grow unevenly causing some bricks to become overloaded before others; therefore, it is often effective to simply add a new node for each overloaded brick.

Resource provisioning is made simpler when the time to bring new resources online is predictable and that resouces can be introduced into the system without significantly disturbing the behavior of existing resources [27]. With its cheap recovery, DStore satisfies both of these assumptions, which further shows that cheap recovery provides a significant point of leverage for experimenting with more techniques like these.

## 8 Discussion and future work

In this section, we wrap up the discussion on administration costs and discuss future work.

### 8.1 Reducing administration costs

Certain kinds of Internet service data, such as billing information, require the transactional semantics and query generality of a relational database. By using a CHT for data that requires durability but can tolerate our relaxed consistency model and does not need to support complex queries, one can reduce the size of the relational database installation, and therefore its administration cost. The overhead in setting up and managing the CHT in addition to an existing database is compensated by the extreme ease of administering the CHT even at large scale.

### 8.2 Future work

One area of future work involves using cheap recovery to evaluate different failure detection techniques and parameters. The low cost of acting on false positives enables us to replace the manual parameter-tuning process with one that discovers more optimal parameters automatically in a similar trial-and-error fashion. A second area of future work involves handling permanent disk failures. With hundreds to thousands of disks, large-scale Internet services replace disks frequently. Rather than requiring immediate replacement, it is important to tolerate a brick being down for several days or more. Once the brick is replaced, the system must automatically integrate the new brick into the system, reconstructing data from the live bricks as necessary.

## 9 Related Work

Motivating our work, Distributed Data Structures [15] is a scalable, high-performance CHT, which uses ROWA and two-phase commit. Session State Management [29], which shares DStore's self-managing goals, is an in-memory CHT that provides non-concurrent-access, semi-persistent storage for session state. Berkeley DB [3], which can serve as underlying storage for DStore bricks, is a single-node hash table that can be replicated using a primary-secondary scheme, but does not provide a single-system image across a cluster.

Brick-based disk storage systems are low-cost alternatives to high-end RAID [7] disk arrays. Federated Array of Bricks [10] uses quorums and a non-locking two-phase protocol to ensure linearizability. To achieve ease of management and performance, RepStore [45] uses self-organizing capabilities of P2P DHTs and a self-tuning mechanism that replicates frequently-written data, but trades write performance for storage by erasure-coding read-mostly data. In contrast to distributed disk and file systems [28, 11, 45, 39] with similar self-management goals, DStore exploits specific workload characteristics for consistency management and exposes a higher-level interface with guarantees on variable-sized elements. Similarly, the Google File System [12] is designed for its large-file, append-mostly workload to achieve scalability and manageability.

Consistency is traded for performance and availability [4, 44], and quorums [13] provide availability under network partitions [8] and Byzantine faults [32, 36]; however, DStore uses quorums and trades consistency for extremely simple persistent state management. Like DStore, Coda [24] tolerates replica inconsistency, but in a non-transparent fashion. Bayou's [38] *Monotonic Reads* and *Read-Your-Writes* provide guarantees similar to those provided by DStore's read-repair and write-in-progress cookie mechanisms; however, Bayou's makes guarantees for a single user session, not across users. The Porcupine Mail Server [37] has self-management goals, but its *eventual consistency* guarantees allow users to see data waver between old and new values before replicas eventually become consistent.

Finally, DStore owes its statistical failure detection to Pinpoint [6], a comprehensive, ongoing investigation of anomaly-based failure detection.

## 10 Conclusion

We combined two techniques—quorums to tolerate replica inconsistency and single-phase operations to avoid locking—to make reboot-based recovery fast and non-intrusive in a cluster-based hash table. One consequence of this cheap recovery is that we successfully applied aggressive statistical anomaly-based failure detection to automatically detect and recover from both fail-stop and fail-stutter transients. Furthermore, we used the same recovery mechanism to solve the distinct problem of incremental scaling without service interruption. The net result is a state store that can be managed with the same types of simple mechanisms and policies as used for stateless frontends.

Taking a larger view, we believe cheap recovery is an important design pattern for self-managing systems: when recovery is predictably fast and has predictably small impact on system availability and performance, the line between "normal" and "recovery" operation becomes blurred. This makes it acceptable for the system to be constantly "recovering," a key property of self-managing, autonomic systems.

## A  Appendix: Algorithm pseudocode

```
put(key, value, ts)
1  if (ts == NULL)
2     ts = generate_timestamp()
3  foreach b in <replica group bricks>
4     send(b, WRITE, key, val, ts)
5  for (c = 0 to WT-1)
6     err = receive(WT_RESP, WT_TIMEOUT)
7     if (err == TIMEOUT)
8        return TIMEOUT

write(key, value, ts)
1  current_ts = read_ts(key)
2  if (ts > current_ts)
3     WRITE(key, value, ts)
4  else if (ts < current_ts - DELTA)
5     return TIMESTAMP_ERROR
6  return SUCCESS

get(key)
1  rset[0...RT-1] = choose_bricks(RT)
2  send(rset[0], READ_VAL, key)
3  for (i = 1 to RT-1)
4     send(rset[i], READ_TS, key)
5  for (i = 1 to RT-1)
6     err = receive(TS_RESP, ts[i], RD_TIMEOUT)
7     if (err == TIMEOUT)
8        return TIMEOUT
9  err = receive(VAL_RESP, ts[0], value, RD_TIMEOUT)
10 if (err = TIMEOUT)
11    return TIMEOUT
12 return check(key, value, ts[], rset[])

check(key, value, ts[], rset[])
1  {maxts, index} = findmax(ts[])
2  timeouts = 0
3  if (ts[0] != maxts)
4     send(rset[index], READ_VAL, key)
5     err = receive(VAL_RESP, value, maxts, RD_TIMEOUT)
6     if (err == TIMEOUT)
7        return TIMEOUT
8  for (i = 0 to RT-1)
9     if (ts[i] != maxts)
10       send(rset[i], WRITE, key, value, maxts)
11       err = receive(WT_RESP, WT_TIMEOUT)
12       if (err != TIMEOUT)
13          timeouts++
14 if (RT - timeouts < WT)
15    put(key, value, maxts)
16 return value
```